# Software architecture and manual for novel versatile CT image analysis toolbox AnatomyArchive

Lei Xu[1,2], Torkel B Brismar[1,2]

[1] Unit of Radiology, Department of Clinical Science, Intervention and Technology (CLINTEC), Karolinska Institutet, Stockholm, Sweden

2. Department of Radiology, Karolinska University Hospital Huddinge, Stockholm, Sweden

Corresponding author:

Lei Xu, email: lei.xu@ki.se

**Abstract**

We have developed a novel CT image analysis package named AnatomyArchive, built on top of the recent full body segmentation model TotalSegmentator. It provides automatic target volume selection and deselection capabilities according to user-configured anatomies for volumetric upper- and lower-bounds. It has a knowledge graph-based and time efficient tool for anatomy segmentation mask management and medical image database maintenance. AnatomyArchive enables automatic body volume cropping, as well as automatic arm-detection and exclusion, for more precise body composition analysis in both 2D and 3D formats. It provides robust voxel-based radiomic feature extraction, feature visualization, and an integrated toolchain for statistical tests and analysis. A python-based GPU-accelerated nearly photo-realistic segmentation-integrated composite cinematic rendering is also included. We present here its software architecture design, illustrate its workflow and working principle of algorithms as well provide a few examples on how the software can be used to assist development of modern machine learning models. Open-source codes will be released at https://github.com/lxu-medai/AnatomyArchive for only research and educational purposes.

## 1 Introduction

To facilitate general-purpose medical image-related clinical studies, it is vital to be able to automatically select or deselect image data with minimum human interventions. It is also greatly beneficial to be able to arbitrarily define and segment the Volumes of Interests (VOIs) in order to extract critical information or features from images, and then to visualize the images and features as well as the intermediate outputs for monitoring or control inspection purposes.

The development of multi-organ or even whole-body segmentation models [1], [2], [3], [4], [5], [6], [7] has laid an important foundation for clinical studies. Usually a laborious effort is required to prepare a dedicated dataset for a clinical study. This is especially true when there have been large variations in scanning parameters and covered body regions between the image acquisitions. When performing retrospective or opportunistic analyses, i.e when the images were acquired for other purposes, confirmation that full coverage of target regions has been obtained is essential. Without automatic solutions, manual inspection and data management would be necessary for every single image data.

After image segmentation, the segmented anatomies are typically organized as a simple collection of objects. Each segmented VOI is usually represented by a unique integer stored in a segmentation image array as in commonly used NifTI image format. This mode of storage does not describe the inter-relationships of anatomies. This makes it unsuitable when analyzing VOIs across multiple hierarchical levels, for example when evaluating the skeleton to determine the risk of fractures, the analysis of either left or right femur, or both, or at the whole skeletal level, or arbitrarily defined multiple associated groups such as specific vertebrae. Moreover, those integers can represent totally different anatomies across different segmentation models. This adds complexity to image database maintenance when upgraded to include new anatomies. The previously stored one-to-one mapping of integer labels to the represented anatomies may then need to be reorganized, breaking previous key-value correspondence.

Measurement of spatial variations within VOIs can provide additional predictive and prognostic information of diseases [8]. This can be provided by voxel-based radiomic features (VbRFs) but not by much more widely used basic segment-based radiomic features (SbRFs). Feature robustness is a critical aspect for training machine learning (ML) models [9], [10], [11]. The feature robustness issue can arise due to intrinsic inter-sample variation [11] and is further compounded by variations in feature extraction parameters [10], [11]. Bin width setting and more critical kernel radius are two influential parameters for VbRF robustness [10]. However, no solution has been proposed to improve VbRF robustness.

Nearly photo-realistic cinematic volume rendering is useful for educational purposes, presurgical and pre-interventional planning [12]. However, there is no existing Python package that provides segmentation-integrated cinematic rendering tool to medical educators and researchers.

To close these gaps, we introduce our open-source package, AnatomyArchive (github repository TBA). The unique functions provided by AnatomyArchive are: 1) anatomy hierarchy- embedded dictionary and multi-graph representation for data storage and anatomy segmentation mask organization, allowing not only flexible VOI definition but also fast data retrieval as well as easy medical image database maintenance; 2) automatic volume standardization according to user's configuration, image data inclusion and exclusion management together with data visualization tools for process monitoring and data control inspections; 3) robust VbRF extraction to better utilize heterogeneity information for training accurate ML models, avoiding unnecessary prefiltering of features for robustness concern at the pretraining phase; 4) GPU-accelerated cinematic volume rendering in Python, allowing segmentation-integrated single and multi-volume visualization. To our humble knowledge, there has not been any computational package addressing these technical needs, neither commercially nor openly.

## 2      Methods

### 2.1      Software architecture and workflow

The architecture of AnatomyArchive is illustrated in Figure 1. It can be described by 10 components, namely the workflowConfig, genericImageIO, segModel, volStandardizer, datasetManager, segManager, featureAnalyzer, simpleStats, simpleGeometry and dataVisualizer. The workflowConfig differs from all the other components, and it is merely a dictionary object with a pre-determined structure. It provides instructions for all tasks and specifies how the tasks should be executed. In the current released code version, it is used as an umbrella term for any dictionary-based input variable container. For future updates, a universally defined workflowConfig variable of nested dictionary type will be constructed to control the entire workflow of the analysis pipeline.

All model-based segmentation tasks are done through the integrated open-source and weights TotalSegmentator [7], allowing segmentations of more than 100 types of organs and tissues in total. It should be stressed that the TotalSegmentator code implementations of versions 1 and 2 as well as their model weights released by the original authors are separately maintained. To ease the functional calls and code maintenance, we have refactorized the codes for configuring models for different segmentation tasks in a way that model weights of both versions can be run by calling the same functions. Version-dependent differences due to reorganization of segmentation classes and associated class maps in similar tasks are handled internally. The corresponding details are covered in subsection 2.3.

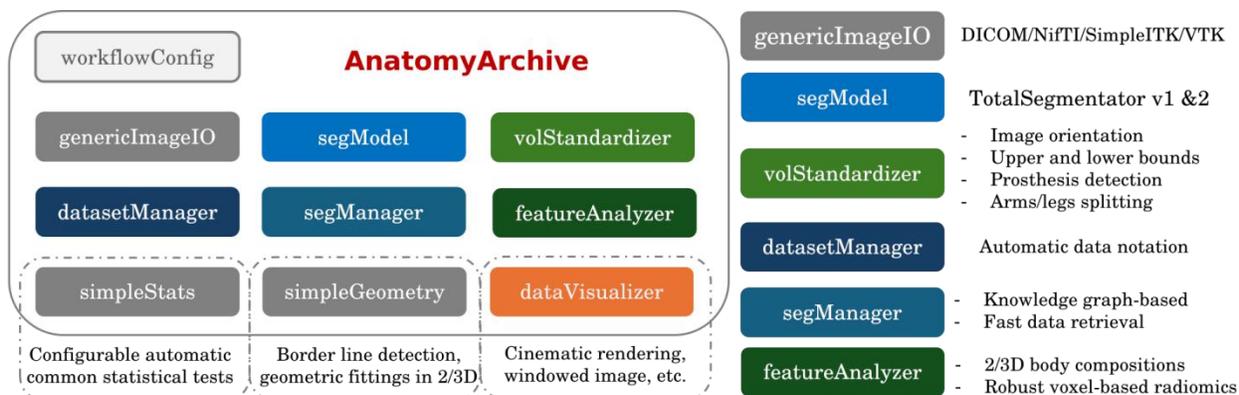

**Figure** 1: Illustration of AnatomyArchive's architecture and summaries of major functions contained in each module except the workflowConfig. The workflowConfig is merely a dictionary object specifying the intended tasks and how they should be executed. It should also be stressed that segModel, which takes TotalSegmentator as an integrated component, is written to be compatible with both versions of 1.5.7 and 2.8.0.

A typical workflow of AnatomyArchive is depicted in Figure 2 below.

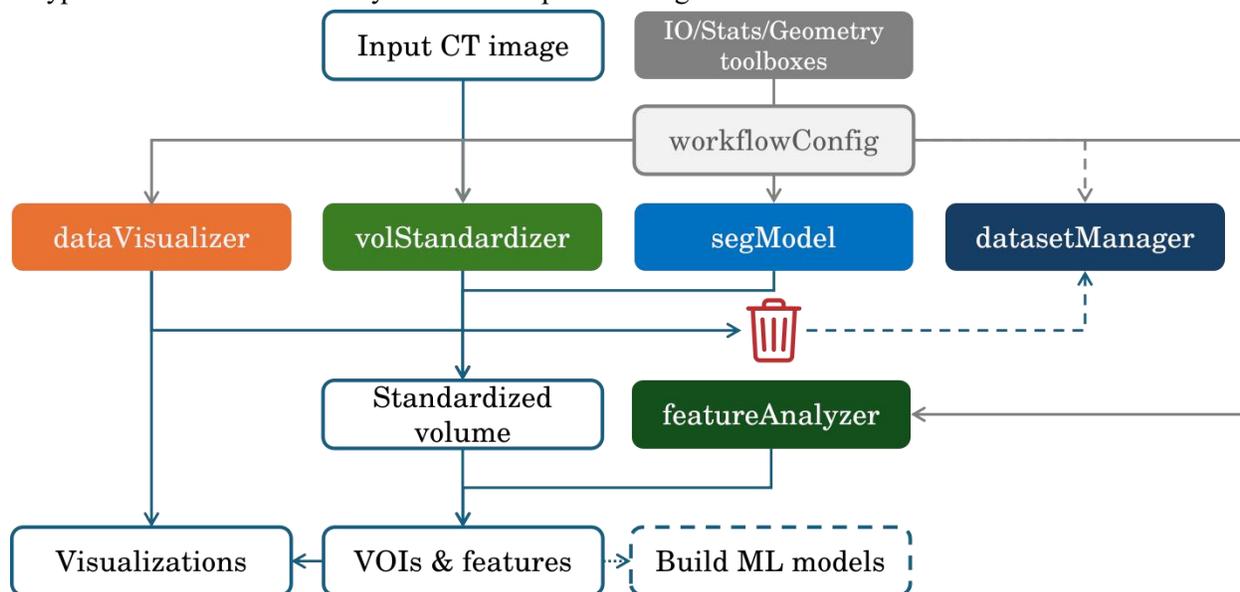

**Figure** 2: Illustrative workflow chart for AnatomyArchive. As workflowConfig controls the entire workflow, all essential components provided by AnatomyArchive are connected to workflowConfig. For the simplicity of the illustrative plot, the GenericImageIO, SimpleStats and SimpleGeometry are clumped together in one modular box and linked with workflowConfig without arrow. They are often recruited to tasks assigned for volStandardizer, segModel, dataVisualizer and featureAnalyzer as indispensable auxiliary components. Their connections to workflowConfig are displayed in gray, to be separated from downstream tasks. Arrows are used to denote information flow. The use of datasetManager is optional, therefore its connections are all shown in dashed lines. AnatomyArchive does not provide dedicated toolsets for training machine learning models. Therefore, this downstream task is also presented using a dashed box and its connection to the selected VOIs and features is also depicted in a dashed arrow. It should be noted that the name workflowConfig referred to here in the current implementation of the released codes is mealy an umbrella term for a dictionary-based configurable multi-variable container. Therefore, it may represent differently structured dictionaries for different connected modules in the workflow chart. For future updated release, a common nested dictionary will be used instead, and only a subpart of it will be used for a specific process.

As the workflowConfig controls the workflow for image inputs, for the executions of intended tasks, information is extracted from it to guide the executions of other modules. For example, from workflowConfig, the volume bounds represented by the upper- and lower-bound reference objects, will instruct the volStandardizer to extract masks of corresponding reference objects via segModel and to compute the bounding planes for each individual input image. The bound-defining objects are supposed to be bone structures. For volume standardization, the workflowConfig may also contain information regarding whether prosthesis or other objects with high attenuation values, presumably metal objects, are to be detected and whether such data should be excluded from downstream tasks or simply notated. Whether body composition analysis is to be performed can be also specified in workflowConfig. If enabled, arm detection will be automatically enabled, and this ensures that segmented skeletal muscles will not contain muscles from the arms in case arms are present in acquired images. This is crucial for quantitative body composition analysis provided by module featureAnalyzer, as there may be patients who have difficulty raising their arms during image acquisitions. For body composition analysis, users may prefer working with values in the conventional 2D plane, or alternatively in a 3D volume specified by the volume bounds.

## 2.2 Package dependencies and versions

The list of dependent packages and recommended as well as required versions is summarized in Table 1.

**Table** 1: list of dependent packages, recommended, and required versions.

| Package name | Recommended version (required) |
| --- | --- |
| TotalSegmentator | 1.5.7 / 2.8.0 |
| python | 3.8 (>=3.5) / 3.10 (>=3.9) |
| torch | 1.13.1 (>1.10.0) / 2.6.0 (>=2.1.2, <=2.6.0; <2.6.0 if python<3.10) |
| dicom2nifti | 2.6.0* / 2.6.2 |
| nnunet | 1.6.6 / N.A. |
| nnunet-customized | ==1.2 / N.A. |
| nnunetv2 | N.A. / 2.6.2 (>=2.2.1, <2.6.0 if python<3.10) |
| numpy | 1.24.4 (<2, >=1.17.3) /- (>=1.19.5) |
| opencv-python | 4.11.0.86 |
| scipy | 1.10.1 (>=1.5.0) / 1.13.1 (>=1.6.0, <=1.31.1 if python <3.10) |
| requests | ==2.27.1 / LCV (==2.27.1 if python<3.10) |
| simpleitk | 2.3.1 / 2.5.2 |
| batchgenerators | ==0.21 / (>=0.23) |
| nibabel | 5.2.1 (>=2.3.0) / 5.3.2 (>=3.2.2) |
| tqdm | 4.67.1 (>=4.45.0) |
| tifffile | 2023.7.10 (<=2023.7.10, >=2019.7.20)/ 2025.5.10 (<=2025.5.10, >=2020.5.25) |
| scikit-learn | 1.3.2 (>=0.14, <=1.3.2) / (>=1.6.1) |
| p_tqdm | 1.4.2 |
| pandas | 2.0.3 (>=0.25.2, <=2.0.3)/ 2.3.0 (>=1.1.3) |
| pygraphviz | ==1.7 / 1.14 (>=1.7) |
| PyOpenGL | 3.1.9 |
| PyOpenGL-accelerate | 3.1.9 |
| PyQt5 | 5.15.11 |
| pyradiomics | 3.1.0 |
| sparse | 0.15.5 (>=0.10.0, <=0.15.5) / 0.17.0 (>=0.14.0, <=0.15.5 if python<3.10) |
| networkx | 3.1 (>=2.5.1, <=3.1) / 3.4.2 (>=2.6.2, <=3.2.1 if python<3.10) |
| matplotlib | 3.7.5 (>=3.3.4, <=3.7.5) / 3.10.3 (>=3.4.0, <=3.9.4 if python<3.10) |
| MedPy | 0.5.2 |
| msgpack | 1.1.1 |
| msgpack-numpy | 0.4.8 |
| trimesh | 4.6.12 |
| vispy | 0.14.2 (>=0.13.0) / 0.15.2 (>=0.14.0) |
| vtk | 9.3.1 / 9.4.2 |

Note: The required versions are provided in brackets. Package versions compatible with TotalSegmentator v.1.5.7 and v.2.8.0 are separated by "/". N.A. means not applicable. If single entries are provided, it represents the latest version, which also works with both versions.

## 2.3 Configurations of pretrained models for segmentation tasks

In the TotalSegmentator package released by the original authors, separately written nested if-statements are used for versions 1 and 2 as in the function totalsegmentator() from the python_api module. In AnatomyArchive, the segModel module instead uses a nested dictionary named segmentation_settings to configure settings to load different models for the corresponding tasks. The key names are designed to reflect the version dependence, using version annotation strings like "v1" and "v2" separated by an underscore symbol after key string "task_id", "crop" or "trainer" in case version dependent settings are necessary. It should be noted that the string "default" is often used as the value string corresponding to the key "trainer" instead of longer string like "nnUNetTrainerV2_ep8000_nomirror" or "nnUNetTrainer_4000epochs_NoMirroring", etc., which is part of subfolder name where the actual model weights for a given segmentation task are located. The "default" string in fact refers to string "nnUNetTrainerV2" and "nnUNetTrainer" respectively depending on the TotalSegmentator version number. AnatomyArchive provides a function from the segModel module `get_seg_config_by_task_name()` to get the version-dependent settings which internally uses the global variable named "v_totalsegmentator" set by the function `get_segmentator_version()`. Please note that the function `totalsegmentator.__version__` returns None. Version number retrieval is done by calling `importlib.metadata.distribution('totalsegmentator').version` after importing the importlib package instead.

In the original implementations of TotalSegmentator for versions 1 and 2, there are separate sets of model weights for segmentation task "total" and "body" depending on the voxel size or the resample variable in python_api. For example, in v2.8.0, three sets of model weights associated with task_id of [291, 292, 293, 294, 295], 297 and 298 are provided for the task "total", each trained respectively with a resolution of 1.5, 3.0 and 6.0mm voxel size. In v.1.5.7, only 1.5 and 3.0mm resolution model weights are provided. Considering that 6.0mm resolution could be too blurry for small lesions, it is decided that only model weights trained with 1.5 and 3.0mm voxel resize resolution are supported in AnatomyArchive. Instead of using a binary flag for variable `fast`, in AnatomyArchive, key strings of "coarse" and "fine" referring to the resolution are used for managing resolution-dependent settings. An example of configuration of version-dependent settings of 1.5mm resolution model is given in the text box below.

```
'fine': {
        'task_id_v1': [251, 252, 253, 254, 255],
        'task_id_v2': [291, 292, 293, 294, 295],
        'task_name': 'total',
        'trainer_v1': 'nnUNetTrainerV2_ep4000_nomirror',
        'trainer_v2': 'nnUNetTrainerNoMirroring',
        'voxel_size': 1.5,
        'crop': None
    }
```

In addition, it is also important to point out that TotalSegmentator has restructured the organization of some segmentation tasks. For example, there are no corresponding model weights to task "bone_tissue_test" of version 1 for the installed TotalSegmentator of version 2. Instead, it can be considered as it has been split into tasks of "appendicular_bones" and "tissue_types" instead. In AnatomyArchive's implementation, the naming of tasks follows the convention of version 2. As a result, for the task of "appendicular_bones" or "tissue_types", if the installed TotalSegmentator version is determined to be of version 1, calling the function `get_v_dependent_cls_map()` will redirect it to the task name of "bone_tissue_test".

Another feature unique about the function `get_v_dependent_cls_map()` is that by default it appends the dictionary defined in the auxiliary class map to its associated main class map if an auxiliary class map exists. Appending them to the main associated class maps provides the possibility to retrieve those anatomy segmentation masks for later use. The concept of an auxiliary class map is introduced in TotalSegmentator version 2. In the original implementation of function `nnUNet_predict_image()` from the nnunet module in TotalSegmetator, performing segmentation tasks that have associated auxiliary class maps, those anatomies defined in the auxiliary class maps will be automatically removed. However, segmentation masks for those anatomies can be valuable for certain applications. Instead, its implementation in the segModel module from AnatomyArchive makes it optional, allowing `remove_auxiliary` as an input variable. Only setting it to be True explicitly will execute automatic removal of auxiliary segments `remove_auxiliary_labels()` function inside our implementation of `nnUNet_predict_image()`. By default, it is set to None.

2.4     Management of anatomies and segmentation masks

For better management of segmentation masks, the segManager module from AnatomyArchive provides pre-defined anatomy groups in a nested dictionary anatomy_groups with key strings of "bone", "digestive_accessory", "intestine", "muscle", "endocrine", "parenchyma", "vasculature", "urinary", and their combinations or extra anatomies to form groups of "cardiovascular", "musculoskeletal", "gastrointestinal" and "digestive". For body composition analysis especially when it comes to skeletal muscles, key words like "iliopsoas", "erector", "gluteus" and "muscle" are used to search for muscles that could be individually analyzed. It shall be noted that sometimes the terminologies used in TotalSegmentator for naming anatomies could either be confusing, for example, "hip" instead of "pelvic" bone, or very rare, for example "autochthon" which is much more commonly referred to as spinal erectors. For anatomy name standardization, a synonym dictionary is defined so that both naming conventions are supported in AnatomyArchive.

As previously mentioned, using integers to present segmented anatomies in an array introduces risks for breaking the one-to-one correspondence if segmentation model is upgraded to include new antomies, as well as complicates cross-hierarchy definition of VOIs. The segManager module from AnatomyArchive provides unique functions to use anatomy hierarchical representation and solves the problem brought by integer-to-anatomy representation. The hierarchical memberships of individual segments are encoded by a directed graph representation, implemented using NetworkX package. It can be converted from a nested dictionary that stores segmented masks by its inbuilt hierarchy. Alternatively, it can be appended to a flat dictionary as an extra component to the stored masks.

To save segmentation masks in dictionaries, AnatomyArchive uses both space and time efficient binary serialization MessagePack package [13], which is widely used in industry 5.0 applications [24]. Instead of saving each individual mask array, which is of the same size as the original image array, coordinate format (COO) in sparse array representation can be much smaller with size of 3J, than original size of M×N×K. This is done via Python library Sparse package, which internally uses standard NumPy and Scipy.sparse Python packages. As MessagePack does not naturally support NumPy arrays, patching function provided by mspack_numpy package is used to pack and unpack 3D mask coordinate arrays. To restore 3D binary mask from coordinates, additional information about the original image size (a tuple of three elements) is needed. In case the segmented anatomy of the original Hounsfield Unit (HU) values is to be restored, either the gray values (of size J) at the 3D coordinates should be saved in addition to the minimum HU gray value as the fill value, or alternatively the user can choose to save the original image data array directly in the dictionary together with all segmentation masks. Multiple memberships of one single anatomy to different groups can be represented by group name-tagged parallel edges that connect it using a multi-graph representation. As such, groups are expressed by the lists of the highest possible hierarchical structures of the constituents without any need to explicitly name all members. For data self-containment, the original image data array can be saved together with masks with or without automatic CT bed removal. Given that MessagePack has not been applied for medical image data storage except AnatomyArchive, most of functions are implemented independently from it.

As CT relies on window settings for better visualization of anatomies, AnatomyArchive groups all segments into three major categories, namely bone, lung and soft tissues (excluding lungs), as shown in Figure 3. This makes it convenient to apply window settings for data visualizations automatically for the concerned VOIs. To demonstrate flexibility in defining multiple cross-organ associated groups, cardiovascular and musculoskeletal systems are included, with their members denoted by colored parallel edges. Meanwhile, left and right anatomical structures are also color-coded to signify that targeted side selection is supported.

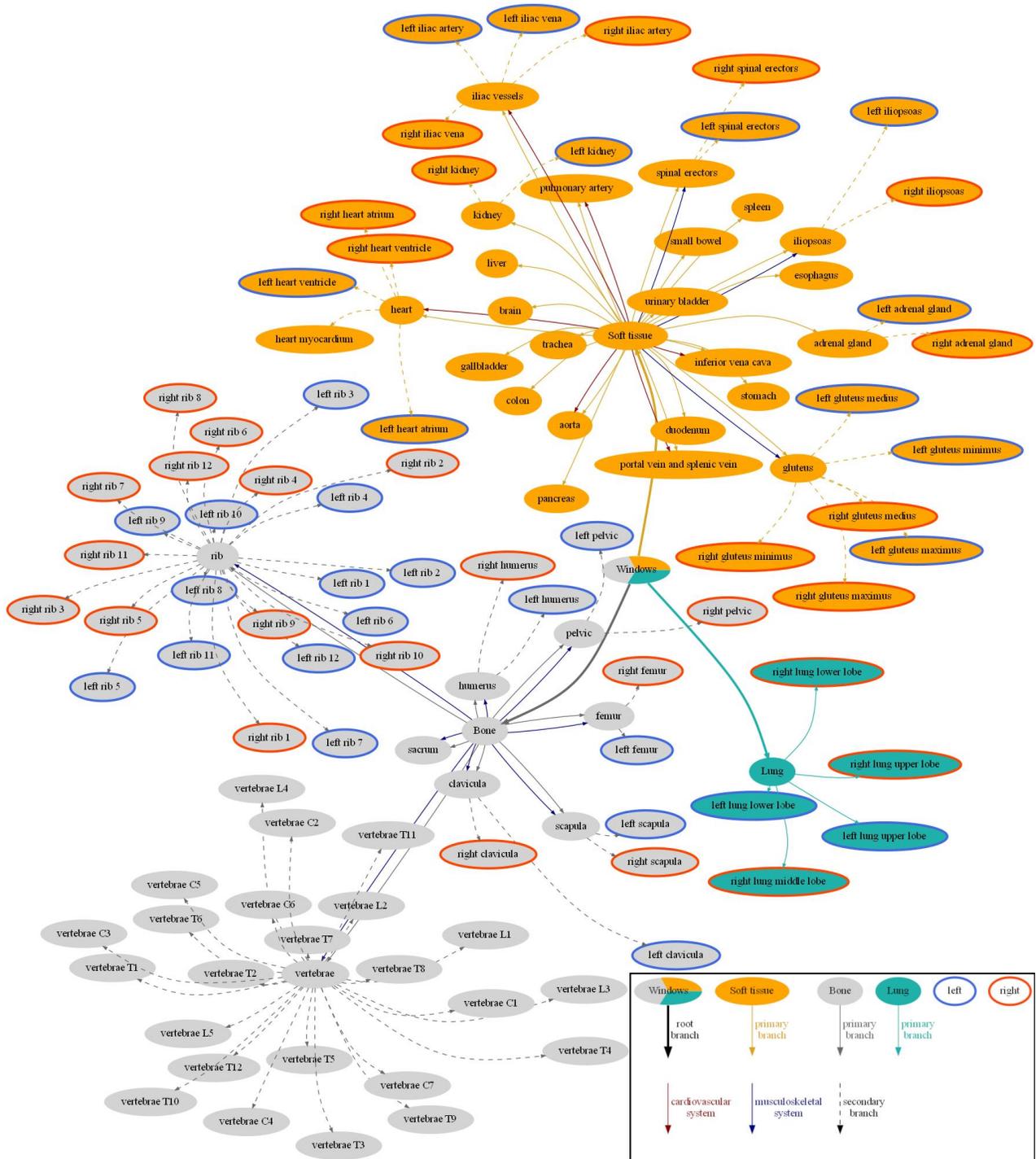

**Figure** 3. A directed multi-graph representation for segmented anatomies with embedded hierarchy. All anatomies are put into 3 major categories associated with applicable CT window settings. The insert shows the color-coding in the graph representation.

## 2.5 Volume standardization and automatic data selection

To automatically select or deselect image data according to users' requirements for the imaged volume, a set of segmentation-dependent functions are provided in the volStandardizer module to standardize the imaged volume and set exclusion reasoning tags for unqualified data.

### 2.5.1 Image orientation standardization

All functions provided by the volStandardizer module require that the images have been pre-reoriented to the same orientation before further processing. The essential image reorientation function is `image_reorientation()`. Its implementation is based on the python package nibabel. However, instead of taking the RAS+ canonical orientation designed primarily for brain visualization (standing side view) as the standard orientation in nibabel package, for the convenience of generic human anatomy visualization, PLS+ orientation (standing front view) is taken as the default in AnatomyArchive. Nonetheless, any arbitrary orientation aligned with as the axcodes specified by the combination of 3-axis denoting characters 'L' (left),'R' (right), 'P' (posterior),'A' (anterior), 'I' (inferior) or 'S' (superior) can be taken as the target or standard orientation as provided by the `image_reorientation()` function. The "+" sign is used to explicitly denote the positive direction of the coordinate system. Only orientation different from the target one will be reoriented. Internally, this is achieved as shown in the code block below, where `io_orientation()`, `ornt_transform()`, `apply_orientation()`, `axcodes2ornt()` are all functions imported from orientations module of nibabel package, while data_arr and affine are respectively the 3D image data array and 4x4 affine matrix.

```
ornt_new = io_orientation(affine) if target_axcodes == ('R', 'A', 'S') else \
    ornt_transform(io_orientation(affine), axcodes2ornt(target_axcodes))
data_arr_new = apply_orientation(data_arr, ornt_new).astype(dtype)
affine_new = affine.dot(inv_ornt_aff(ornt_new, data_arr.shape))
```

### 2.5.2 Volume bounds

The volStandardizer module of AnatomyArchive packages provides a bound-defining function `define_volume_bounds_by_anatomies()` using dictionary-based user-configured upper- and lower-bound anatomies after image orientation standardization. Typically, bones are recommended to be used as bound-defining anatomical structures. In case bound-defining anatomies are cropped or entirely missing, based on the segmentation results provided by the integrated TotalSegmentator, negative integer values of –1 and –2 respectively will be set in the bound-defining dictionary. If the bound-defining anatomies are in fact entirely covered, the computed indices of upper- and lower-bound planes in the z-axis will be set in the bound-defining dictionary. The bound planes will then be used for the entire workflow. It is worth mentioning that as there exist anatomies in human which have both left and right counterparts, it is unnecessary to specify that the upper- or –lower-bound defining anatomy must be the left or the right one in case the bound should be defined all together by them. In this case, the bound-defining anatomy can be set using the common string without left or right annotation. However, if there are more than two anatomies sharing the same common string used for denoting the bound-defining object, an error will be raised.

It is also recommended to use a nested dictionary-based variable dataset_tag for process control together with variable data_id which is image specific. In case negative integer values for volume bounds are obtained, errors in bound-defining anatomies will be automatically annotated and used to instruct the skipping of downstream tasks. In the meantime, the data_id will be appended to a list of the nested dictionary dataset_tag taking the associated error code as the key string. This functionality helps the user easily identify the image data excluded from downstream tasks and verify that this is done correctly.

The algorithm for bound detection is depicted in Figure 3.

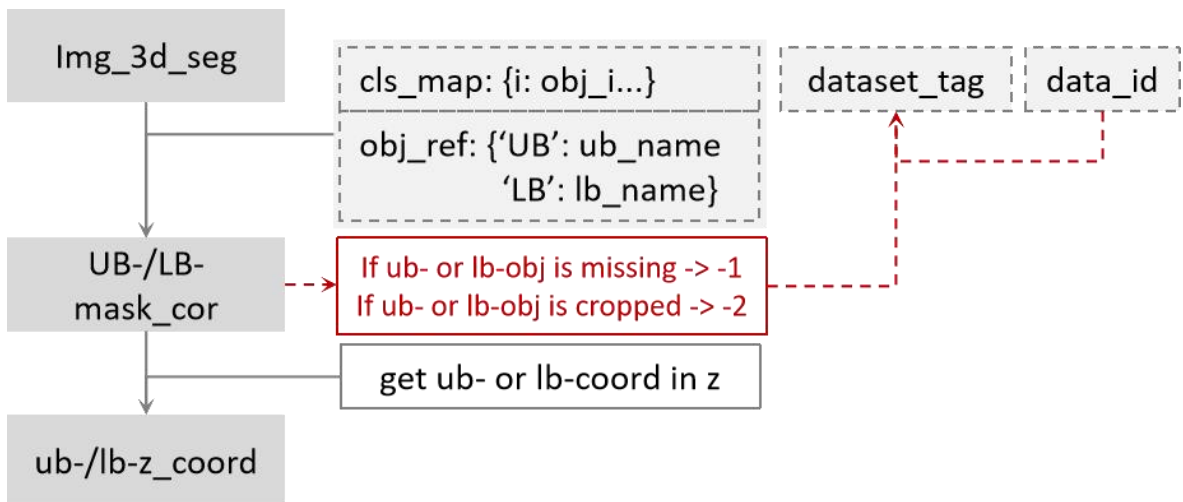

**Figure** 3: the illustration of the upper- and lower-bound detection algorithm. The function `define_volume_bounds_by_objects()` takes class map name to load the corresponding class map `cls_map` to identify the masks of bound-defining objects which are specified in the input dictionary `obj_ref`. Thereafter, the upper- and lower-bound mask projected to the coronal plane are used to compute their upper- and lower-bound planes respectively for downstream processes. In case the bound defining object is missing or cropped in the segmentation results, negative integer values are returned in the intermediate process and that information can be added together with the `data_id` to the dictionary-based variable `dataset_tag`.

The cropping detection is based on checking if a 3D mask projected to its transverse, coronal, and sagittal planes using Maximum Intensity Projection (MIP) touches any side of the boundaries. If the segmentation mask has been deliberately segregated from the boundaries by adding margins via variable `crop_addon` when performing segmentation via function `nnUNet_predict_image()`, these offsets can be also applied.

### 2.5.3 Hip prosthesis detection

Prosthesis detection enables automatic annotation of image data in case images with prosthesis need special processing or in case they should be excluded to fulfil the need for the aimed study. TotalSegmentator offers nnUNet model weights to segment prosthesis in CT images, which can be easily used for prosthesis detection in case the segmentation mask exceeds a size limit. Alternatively, it is also possible to detect hip prosthesis based on the notion that they are made of metals, which have higher attenuation values than bones and almost always intersect the cross-sectional plane of the lower end of pubic bones. Therefore, if pubic bones are set as the lower-bound reference object, the hip prosthesis can be accurately detected by identifying bright objects that exceed a size limit and simultaneously intersect with the lower bound. The bright object segmentation function `segment_bright_objects()` is based on edge detection using the powerful Farid filter[14] combined with watershed algorithm [15] and morphological operations.

The algorithm for lower-bound prosthesis segmentation is illustrated in Figure 3.

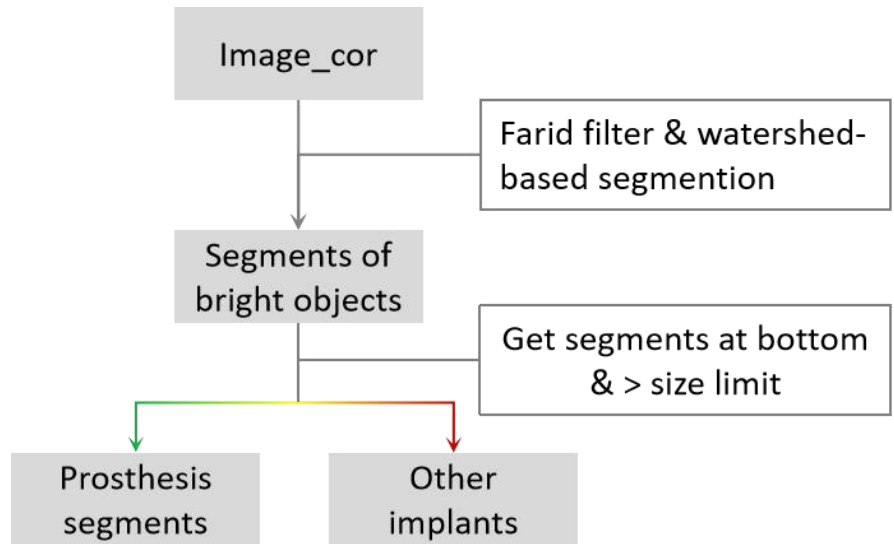

**Figure** 4: algorithm to segment prosthesis from coronal plane image after maximum intensity projection. Function `segment_bright_objects()` based on Farid filter, watershed algorithm and morphological operations is used to segment bright objects from the coronal plane image. Among the segmentation results, objects which are at the image bottom and exceed a size limit are considered prosthesis segments, otherwise other implants.

Python pseudocodes for hip prosthesis detection are shown below.

```
img_cor_cropped = np.flipud(np.max(img_3d[:, :, lower_bound:], axis=0).T)
seg_labeled, seg_prosthesis = get_lb_bright_objects(img_cor_cropped)
prosthesis_detected = 0 if np.isscalar(seg_prosthesis) else 1
if prosthesis_detected:
    if dataset_tag is not None:
        add_tag_to_data(dataset_tag, 'prosthesisDetected', data_id, 'Warning')
```

### 2.5.4  Arm detection and segmentation

The detection of arm presence and segmentation in the image is especially useful for body composition analysis such that unexpectedly present arms can be removed from the analysis. This is enabled upon integrating the body segmentation task provided by TotalSegmentator and using arm-identification algorithms uniquely provided by AnatomyArchive. The mask labels returned by TotalSegmentator do not differentiate arms from legs by using the same integer label of 2 representing the body_extremities apart from the mask for `body_trunc` with integer label of 1. Function label() from the measure module of skimage package is used to split all body extremities into different entities. As only arm-associated bones will overlap with the arms, function `separate_arms_and_legs()` uses arm-associated bone masks of humerus, ulna and radius to separate arms from legs, given the segmented extremities. TotalSegmentator of version 1 offers a model for the task "bones_tissue_test" which provides the full set of humerus, ulna and radius to help identify arms. While for version 2 of TotalSegmentator, humerus is only present in class map "appendicular_bones_auxiliary", which is associated with task "appendicular_bones". As previously mentioned in subsection 2.3, any anatomy in an auxiliary class map, like humerus in this case, will be automatically removed after executing the `nnUNet_predict_image()` function implemented in the original TotalSegmentator package. This can create a problem for the case where only upper arms but not forearms are covered in the image. This is not the case as introduced in subsection 2.3 for AnatomyArchive's implementation.

The function of `separate_arms_and_legs()` can be also used for body cropping detection. It is based on the segment cropping detection algorithm introduced in subsection 2.5.2 and takes the dictionary of upper and lower bounds also as an input variable to determine whether the body trunk within the upper- and lower-bounds are fully covered in the scanned image. If the body trunk within bounds is cropped, the computed volumes of body compositions like skeletal muscles and fats will be inherently inaccurate. Therefore, these cases must be excluded from the dataset if body compositions are an essential part of features for downstream ML model building.

## 2.6 Feature analysis

All feature analysis functions are provided in module featureAnalyzer. It includes functions for body component analysis and radiomic feature extraction, specifically targeting voxel-based features.

### 2.6.1 Body component analysis in 2D and 3D

Body component analysis refers to generic analysis of segmented anatomies, like organs or tissues. The body *composition* analysis is considered a subset of body component analysis. It refers to primarily analysis of skeletal muscles including muscle quality, different types of fats, and so forth. The function `body_component_analysis()` from module featureAnalyzer allows generic analysis of body components and specific body composition analyses in both 2D and 3D forms.

For execution of the `body_component_analysis()` function, a variable `target_eva_config` of nested dictionary type must be defined. The keys of its first hierarchical level are supposed to be class map names. It is designed in such a way because each class map is associated with a TotalSegmentator segmentation task.

In case the reference object(s) will be used to set upper- and lower-bounds or only to set the central plane, a reference class map will also be needed. By default, the class map name "total" is supposed to be the reference class map. It is designed in this way as it is of the highest probability that the reference object(s) would be contained in the class map, though there is no guaranty that the reference object(s) would be present in the image. Missing reference object(s) error in images can however be handled by `dataset_tag` as mentioned in subsection 2.5.2.

The supported key strings for 1st and 2nd level and corresponding value type for 2nd level keys are provided in Table 2.

Table 2: Typical parameters used in dictionary `target_eva_config` for task-specific settings.

| Type of tasks | 1st level key string | Supported 2nd level key string | 2nd level value type |
|---|---|---|---|
| Regular task | Any supported task | ‡ "selectedObjs" | List of string |
| Reference object-defining task | "total" as default | "refObj" | string |
| | | "refObjUB" | string |
| | | "refObjLB" | string |
| | | "coarse" | Boolean |
| | | "excludeProsthesisSamples" | Boolean |
| | | "enforceMuscleRange" | Boolean |
| | | "dict_hu_range" | dictionary |
| Task for tissue types | * "tissue_types" | "enforceMuscleRange" | Boolean |
| | | "enforceFatRange" | Boolean |
| | | "dict_hu_range" | dictionary |

‡ indicates that the key string is supported by all tasks but not explicitly listed for all. Please note that all other 2nd level key strings are optional and can be omitted.

* indicates that though task name "tissue_types" exists only in version 2 of TotalSegmentator, AnatomyArchive reuses task "bone_tissue_test" for its purpose. More explanation is found in subsection 2.3.

If key strings of "refObjUB" and "refObjLB" are used for reference object-defining task, upper- and lower-bounds will be calculated to standardize imaged volumes. If key string "refObj" is used, the central plane of the reference object will be used to calculate body component attributes in 2D. When there is no key string to denote any reference object(s), no volume bounds or central plane will be calculated. The task is then reduced to a regular one. Key string "coarse" and the corresponding binary value instruct whether a coarse or fine resolution model is used for the segmentation task. By default, the fine resolution model is used. Regarding differences in coarse and fine resolution models, more information is found in subsection 2.3. Besides, the option with "excludeProsthesisSamples" controls whether the prosthesis-containing image data should be excluded from downstream tasks. By default, it is set to False. Other supported key strings including "enforceMuscleRange" and "dict_hu_range" can be provided by the users. For task "total", segmentation of various types of muscles is included, as mentioned in subsection 2.4. If "enforceMuscleRange" is set True, the segmented muscles will be further split into 3 categories and the fat within muscles is known as the inter-muscular adipose tissue. It would be done using either the default `dict_hu_range` given in Table 3 or values defined by the users.

Table 3: Default HU value ranges [16] for different types of tissues defined in AnatomyArchive.

| Tissue types | HU range |
|---|---|
| Normal attenuation muscle | [30, 150] |
| Fat | [-190, 30] |
| Fatty muscle / low attenuation muscle | [-29, 29] |

For task "tissue_types", the key string of "enforceFatRange" is also supported as subcutaneous fat and torso fat are included in the segmentation results. If set to True, the segmented fats will be forced to be within the range defined by `dict_hu_range`.

### 2.6.2 Automatic histogram bin width setting

Doane's bin width optimization algorithm [17] is used to automatically set optimal bin widths for histograms. The function `get_optimal_hist_bin_width()` defined in module simpleStats is implemented internally using numpy function `histogram_bin_edges()`, where the input variable bins= 'doane'. However, instead of using directly the optimized bin width, which can be a float number, the returned value is first rounded to the nearest integer before being further rounded to any value from a user-preferred array. By default, the rounding targets are [2, 5, 10, 20, 40, 50], i.e., 2.3 will be rounded to 2, while 16.8 will be rounded to 20. This rounding option is provided in function `set_type_bin_width()`.

### 2.6.3 Robust voxel-based radiomic feature extraction

For conventional radiomic feature extraction, the featureAnalyzer module from AnatomyArchive implements its own feature extraction function and extractor initialization using featureextractor module from the PyRadiomics package. There are several reasons for this choice. Firstly, it has been discovered that having the variable initValue set to NaN for feature computation in a YAML parameter file can lead to numeric errors later. To handle this issue, it will be automatically replaced by zero if detected. Secondly, because the mathematical definitions of energy and total energy for the first order feature, as well as those of joint average and sum average for gray level co-occurrence matrix differ only by a scaling pre-factor, it is decided to compute only the energy and joint average features. To enable voxel-based feature extraction, a Boolean flag voxel_based variable is set by the user when initializing the feature extractor during loading of the parameter file using function `get_feature_extractor()`.

Besides, function `get_voxel_based_radiomic_features()` provides support for feature extractions for a list of selected integer labels with a list of correspondingly optimized bin widths via variables of `selected_labels` and `preset_bin_widths`. This feature is not supported in the original implementation of PyRadiomics package, for which only single variables of integer label and bin width are used. In principle, it is possible to write the function to be compatible with the VOIs represented using mask coordinate formats stored in MessagePack file as proprosed in AnatomyArchive. This could be done by including an internal file type conversion to, for example, NifTI image file format to use the SimpleITK-dependent featureextractor module. This, however, has not been implemented yet.

To obtain robust VbRF that is advantageous in retrieving intra-segment heterogeneity information for ML task, standardization-based Subset Average Pooling (SAP) has been developed to improve VbRF robustness. SAP is inspired by the average pooling for feature map down-sampling used in conventional neural networks [18]. The major conceptual difference is that the purpose of SAP is not to reduce the size of feature map within a window region, but to extract representative feature vector for a subset of features of the same type extracted under various conditions, for instance, different kernel radii. It is expressed as:

$$X'_i = \frac{X_i}{\log N},$$

$$Z_i = \frac{X'_i - \overline{X'_i}}{\sigma_{X'_i}},$$

$$SAP = \frac{1}{k}\sum_{i \in S}^{k} \frac{Z_i - \overline{Z_S}}{\sigma_{Z_S}}.$$

$X'_i$ represents a VbRF vector $X_i$ extracted in condition $i$ normalized by the number of voxels $N$ within the VOI, similar to the study [20] proposed for correcting voxel-size and volume dependence of SbRFs. $Z_i$ represents the standardized version of $X_i$ using its mean $\overline{X'_i}$ and standard deviation $\sigma_{X'_i}$. And $Z_i$ with $\in$ is standardized again using the mean $\overline{Z_S}$ and standard deviation $\sigma_{Z_S}$ computed from the subset of conditions $S$ of size $k$ and computes their averages to get a feature vector that has the same size as $X_i$ extracted in an arbitrary condition for feature robustness evaluation. It is important to note that users can choose to compute a VbRF in arbitrarily many conditions. Even if one takes the VbRFs computed under all chosen conditions for SAP computation, they are still a subset of all theoretically allowed feature extraction conditions, which is practically bounded by the size of a VOI.

## 2.7 Data visualization

AnatomyArchive offers a set of functions for data visualization, including preconfigured window settings for different cross-section planes with and without masking, generation of control images for inspection according to user-configured data inclusion and exclusion criteria, voxel-based feature visualization, conventional 3D volume rendering as well as GPU-accelerated segmentation-integrated cinematic volume rendering. For image visualization in 2D planes, all visualizations are implemented using the matplotlib package. For 3D image visualizations, depending on the intended visual effect, packages like VisPy or VTK may be used instead.

### 2.7.1 Windowed image

As with many medical image analysis packages, AnatomyArchive also provides window settings for CT images, mainly for visualizing lungs, soft tissues other than lungs and bones. The default window settings are provided as shown in Table 4.

**Table** 4: Default window settings for different anatomical types.

| Anatomical types | Window width | Window level |
|---|---|---|
| Lungs | 1500 | -600 |
| Soft tissues other than lungs | 350 | 50 |
| Bones | 1800 | 400 |

Depending on the configured window width win_width and level win_level, the displayed min and max are calculated as: `disp_min = win_level - win_width/2` and `disp_max = win_level + win_width/2`.

What is unique about AnatomyArchive is that depending on the selected anatomies configured by the users, the window settings will be automatically applied during image visualization. The selection of the correct window settings for image visualization is achieved depending on the types of selected anatomies and the intended tasks.

### 2.7.2 Process control

Image data visualization for process control is one of the key unique features provided by AnatomyArchive. The automatically generated control images include visualization of image data with computed bounds represented by straight lines, visualization of prosthesis in color if detected, cropped body contours with dashed-dot lines in colors, visualization of arms with outlined contours in colors if detected, and central plane of a reference object represented by a straight line. All these plots are made by projecting the 3D image data array to 2D planes via MIP algorithm and are mostly saved in PNG format. In most cases, images are displayed in the coronal plane. However, for body cropping detection, either the image is presented in the coronal plane or the sagittal plane depending on along with which axis the body trunk is found partially covered in the scanned image. The contour visualization is achieved with help of `find_contours()` function imported from the measure module of skimage package.

### 2.7.3 Voxel-based feature visualization

The visualization of VbRFs is nested inside the function `get_voxel_based_radiomic_features()`.

The VbRFs extracted by the featureextractor module from PyRadiomics restrict the output feature maps to a rectangular cuboid bounding box calculated from the original mask using member function `GetBoundingBox()` of LabelStatisticsImageFilter provided by SimpleITK package. As a result, the calculated feature maps of the SimpleITK.Image type are only locally defined with respect to the bounding box of a specific mask and need to be reindexed for global reconstruction. The global reconstruction of the feature maps from the cropped ones is achieved using python pseudocodes shown below.

```
mask_arr, voxel_size = convert_sitk_image_to_numpy(mask_sitk)
mask_binary = mask_arr == _label
x_0, y_0, z_0 = np.min(mask_coord, axis=0)
x_s, y_s, z_s = np.max(mask_coord, axis=0) - np.min(mask_coord, axis=0) + 1
fmap_cropped, _ = convert_sitk_image_to_numpy(fmap_sitk)
fmap = np.zeros(mask_arr.shape)
fmap[x_0:x_0 + x_s, y_0:y_0 + y_s, z_0:z_0 + z_s] = fmap_cropped[ks:-ks, ks:-ks, ks:-ks]
```

Thereafter, a reindexed feature map is normalized and rescaled to 8-bit values and then appended to a 4 channel-RGBA format for later display using Matplotlib.

```
mask_binary = recover_3d_mask_array_from_coord(mask_coord, data_shape)
mask_channels = mask_binary[..., np.newaxis]
fmap_ma = np.ma.masked_array(fmap, mask=~mask_channels)
fmap_ma [..., 0] = (fmap_ma[..., 0] - np.ma.min(fmap_ma[..., 0])) / \
                                    (np.ma.max(fmap_ma[..., 0]) -
np.ma.min(fmap_ma[..., 0])) * 254 + 1
fmap_ma = fmap_ma.filled(fill_value=0)
fmap_ma = np.concatenate((fmap_ma / 255, np.zeros(tuple(list(data_shape) + [2])),
                          1 * mask_binary[..., np.newaxis]), axis=3)
```

### 2.7.4 Cinematic volume rendering

The GPU-accelerated cinematic volume rendering in AnatomyArchive is based on the VTK (Visual Toolkit) package, the same as the underlying package used by open-source software like VolView [19] and C++ based 3D Slicer[20] for the cinematic rendering functionalities. VolView is said to run 100% locally on users' own machine and is supposedly data secure. However, access to its WebAssembly-based rendering function requires internet access. More importantly, for local applications, the choice of web-based programming languages only complicates its deployment for Python-only applications. 3D Slicer is a comprehensive software package for medical image analysis and visualization, and it has integrated TotalSegmentator as well to enable full body segmentation. However, it is too bulky to integrate into a third-party package. In terms of cinematic rendering, one unique feature of the volRenderer module from AnatomyArchive package is that it allows not only the whole volume rendering just as VolView and 3D Slicer, but also segmentation-integrated composite rendering, which has not been demonstrated in published literature. The volRenderer module from AnatomyArchive package offers functions to render each anatomy or anatomies of the same group using a separate set of parameters for volume rendering and visualization in the same view.

Each rendering effect is controlled by key parameters, e.g. diffusivity, specular power, ambient lighting, gradient opacity, scalar opacity, color transfer function, etc. These parameters are user-configurable and can be either manually set or directly loaded from 3D Slicer's preset file in XML format by specifying the string name of the desired rendering if users intend to reuse the rendering settings provided by 3D Slicer. The parsing of XML elements to Python dictionary-based rendering settings is done via function `get_rendering_preset_from_xml()` in volRenderer module which relies on external xml.etree.ElementTree module. The returned dictionary preset_dict variable(s) can then be used to generate different cinematic rendering results via functions like `create_rendering_for_volume_with_preset()` for whole volume uniform rendering, or `create_composite_rendering_with_mask_associated_presets()` for depth-unaware composite rendering or `multi_volume_rendering_with_presets()` for depth-aware segment-integrated multi-volume rendering.

The difference in depth-awareness for composite rendering comes from the difference in the actors for volume rendering (see demos in Figure 7A&B). In the backend, the depth-unaware composite rendering uses the same rendering actor vtk.vtkVolume as for conventional whole volume uniform rendering. The consequence is that the last rendered object, regardless of its relative positioning to others, will always be rendered on top of the scene. It should also be stressed that a distinction is made between foreground and background rendering settings in the composite rendering function. As a result, each individual foreground object is allowed to have its own rendering settings since the input variable `preset_fg` can be a list of dictionaries, while the background objects will always be rendered in the same way using a single dictionary `preset_bg`. This makes it particularly suitable for putting a foreground object into focus in the presence of background object rendering. However, having multiple foreground objects without depth-awareness can cause spatial confusion for users.

This spatial confusion problem is solved via `vtk.vtkMultiVolume` rendering actor which is employed in the `multi_volume_rendering_with_presets()` function. In this function for the input variables, no distinction is made between foreground and background object rendering. Users shall supply a list of rendering settings referred to as variable `list_preset_dict` accompanied by the list of separately saved masked NifTI image files referred to as `list_image_files`.

### 2.8 Statistical tests

AnatomyArchive has integrated commonly used statistical test functions in the simpleStats module. For example, widely used variants of t-tests and f-tests to compare two groups of samples for their means and variances are integrated. Function `ttest_with_auto_checks()` returns a dictionary of p-values, and chosen test names for statistical tests depending on automatic evaluation of data normality, the sensitivity level on normality deviation, whether the sampled groups are paired or independent, the choice of alternative hypothesis, i.e., being "two-sided", "greater" or "less" and whether variants of f-test should be included. The normality test is performed using D'Agnostino and Pearson's test. Depending on whether the conditions are met for each statistical test, regular t-test, Welch's t-test, paired t-test, Mann-Whitney U test or Wilcoxon signed-rank test will be automatically chosen. Similarly, the regular f-test, Levene' test, Brown-Forsythe's test or its trimmed version will be automatically selected for testing variance equality, depending on requirement on normality with regards to the degree of data skewness and kurtosis as well. To our knowledge, this is a unique feature of simpleStats offered by AnatomyArchive for statistical tests.

Other commonly used statistical methods included in the simpleStats module are calculation of intra-class correlation coefficient [21], Overall Concordance Correlation Coefficient (OCCC)[22] for rater reliability evaluation, DeLong's test [23] for comparing Receiver Operating Characteristic (ROC) curves for model performance evaluation, Kolmogorov-Smirnov test which can be both used for comparing histograms and evaluating model performances as an alternative to DeLong's test [24], calculation of the confidence interval of the Area Under the Curve (AUC) value for a ROC curve and estimation of sample sizes for linear mixed models based on power calculations.

Some plotting functions very closely related to statistical measurements are also provided in the simpleStats module, such as kernel density estimation, heatmap visualization and annotation with statistics, evaluation of feature robustness with integrated statistical tests, boxplots with integrated statistical tests, etc.

### 3 User manual and demos for selected functions

### 3.1 Body composition analysis in standardized volume

Suppose there is an application scenario that one would like to use L1 vertebra and pubic bones (the lower end of pelvic bones) to set the upper- and lower-bound respectively to standardize CT volumes, and to obtain volumetric metrics about subcutaneous fat, torso fat, and skeletal muscles without applying the `dict_hu_range`. To fulfill these needs, one needs to define the variable `target_eva_config`.

```
target_eva_config = {'total': {# 'refOb': 'vertebrae_L3', # For 2D analysis
                    'refObjUB': 'vertebrae_L1',
                                        'refObjLB': 'pelvic',
                    'excludeProsthesisSamples': True}
                            'tissue_types': {'selectedObjs': ['subcutaneous_fat',
'torso_fat', 'skeletal_muscle'],
                                            'enforceMuscleRange': False}
                    }
```

It shall be noted that AnatomyArchive works with the task name of "tissue_types" regardless which version of TotalSegmentator is installed on the users' computers as clarified in subsection 2.3.

To perform the desirable analysis via body_component_analysis(), apart from variable `target_eva_config`, one needs to provide the directory of image files `dir_input`, initialize the `result_dict` via `NestedDict()` imported util module. It is highly recommended to initialize automatic process control variable `dataset_tag` via `NestedDict()` as well to keep track of included image data.

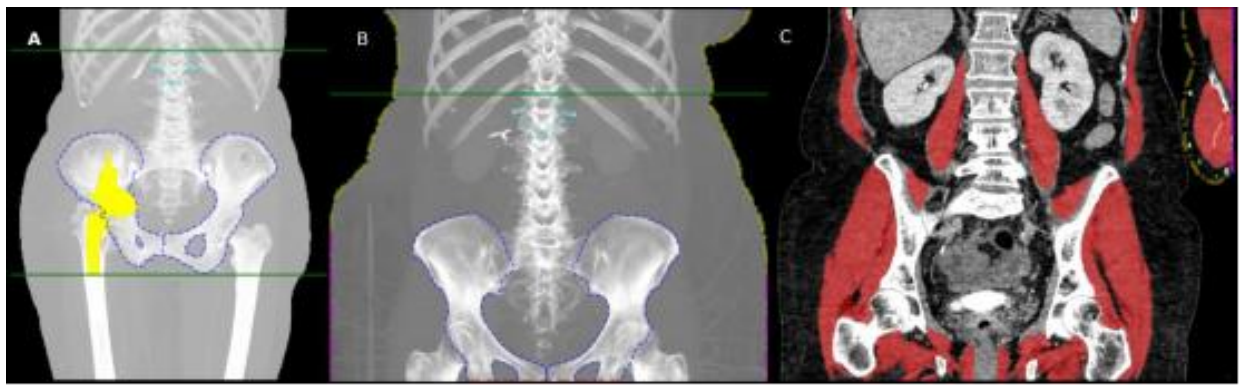

**Figure** 5: Visualization of segmentation-based volume standardization functions. A: User-configurable settings of upper (lumbar vertebrae L1) and lower-bound (pelvis) for volume standardization and automatic edge-based prosthesis detection (yellow) at lower pelvic level, whereas the returning flag on detected prosthesis may be used to instruct exclusion of the patient instance. The coronal image is visualized with automatically set maximum Hounsfield Unit (HU) value excluding any automatically detected bright objects. B: Automatic detection of cropped boundary-defining anatomies and body. Failure to contain intact upper- and lower bound-defining anatomies (dashed red line) triggers automatic exclusion of the image stack instance, while cropped body (magenta dash dot line) on the sides sets a flag that the image stack should not be used for studies involving body composition analysis. The coronal image is shown with an automatically set maximum HU-value to avoid visual effect of over-saturation or being gloomy because of inappropriate maximum setting.

C: Automatic separation of arms (enclosed by yellow and magenta dash dot lines) from body trunk and legs, which is critical for body composition analysis especially skeletal muscles (red shades). The coronal image visualizes an automatically retrieved slice with the maximum cross section area of skeletal muscle after application of the soft tissue window.

### 3.2 Robust VbRF extraction

Suppose that the VOI is the right kidney, i.e., selected_label = 2. To demonstrate the robustness of the VbRF extracted via SAP method introduced in AnatomyArchive in comparison with baseline and conventional feature standardization, experiments can be performed by varying the kernel radius using settings of 2, 3, 4 and 5, therefore `target_param = 'kernel_radius'`, `target_range = [2, 3, 4, 5]`. Set the file path to "exampleVoxel.yaml" file with settings shown as below:

```yaml
imageType:
  Original: {}
featureClass:
  firstorder:
setting:
  binWidth: 25
  force2D: false
  label: 1
voxelSetting:
  kernelRadius: 2
  maskedKernel: true
  initValue: 0
  voxelBatch: 10000
```

Set `data_dir` to a local directory of input images. Intialize `dict_results=util.NestedDict()`. Set variables `do_ttest = True` and `plot_result=True`, set `save_stats_path` to a path on a directory and run function `test_radiomic_feature_robustness()` with all of the specified input variables. The OCCC values for the same types of features extracted in 4 different conditions as the baseline and those after conventional feature standardization will be automatically computed. Thereafter, set `n_components` to 2 or 3, keep the already defined variables as they are, and run function `eval_feature_robustness_with_sap()`. In this case, the SAP approach will be performed by randomly selecting features generated using 2 or 3 different kernel radii to form the subset of testing conditions. And this will resulted in $C_{4,2}=6$ or $C_{4,3}=4$ combinations of sampled subsets and their OCCC values will be computed. The results will then be compared to baseline OCCC values or values after conventional feature standardization.

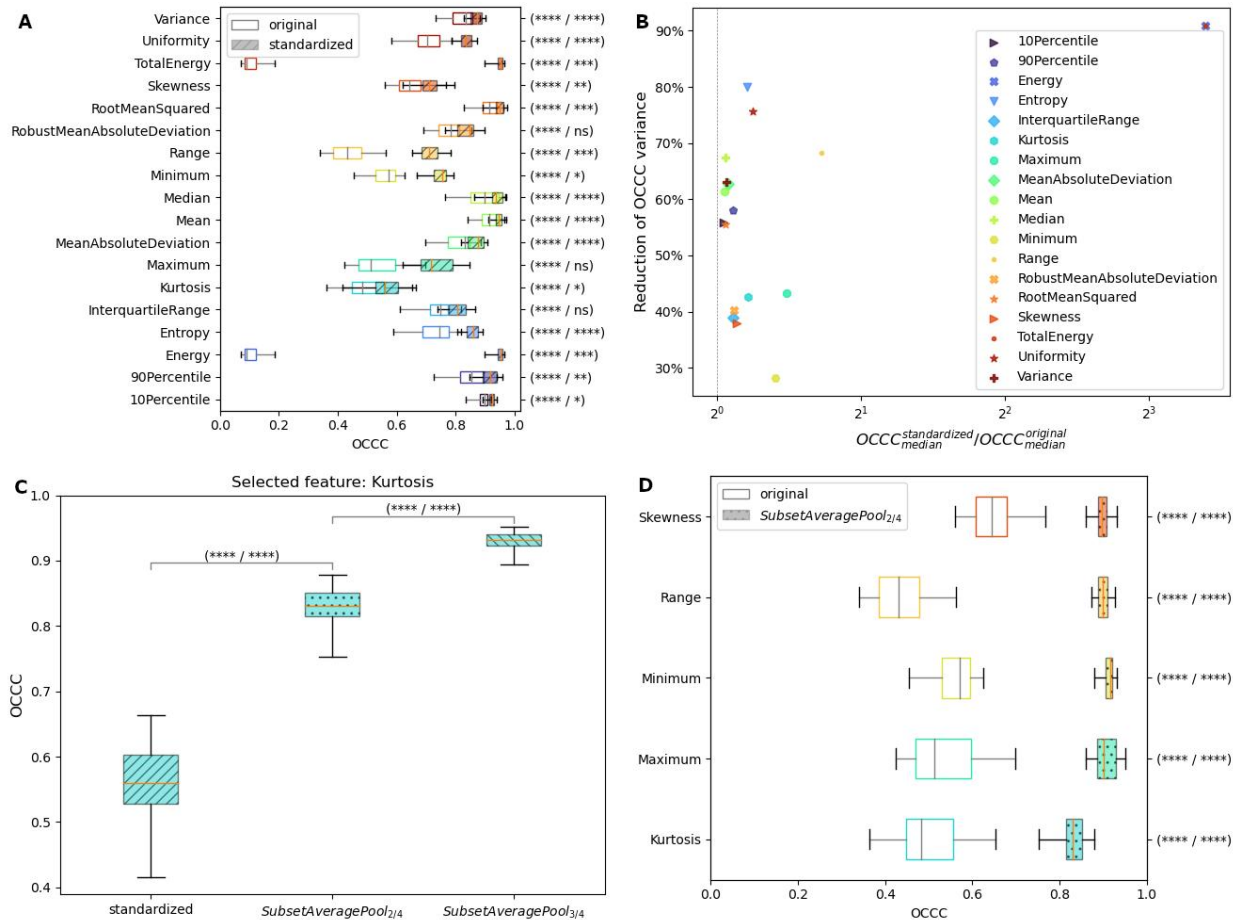

Figure 6. Comparison of feature robustness evaluated by the Overall Concordance Correlation Coefficient (OCCC) value for original first-order voxel-based radiomic features extracted on 4 different kernel radii compared to features after standardization and proposed SAP approach. A:Box plots of all original (empty box) first-order feature versus the standardized (filled and forward slash-hatched box) features. Each box plot shows the 1st, 2nd and 3rd quartiles with extended caps representing the 2.5th to 97.5th percentiles. The second y-axis on the right shows the significance levels for differences in the median and variance value for each respective feature. B: Scatter plot of reduction in OCCC variance in percentage versus the ratio of median OCCC values after standardization divided by that of the original. C: Box plot comparison of kurtosis after standardization (filled and forward slash-hatched) versus SAP with random sampling of 2 (filled and dot-hatched) or 3 (filled and backward slashhatched) kernel radii out of all 4 conditions. D: Box plot comparison on OCCC values of selected original features (empty box) versus features after SAP with random sampling of 2 (filled and dot-hatched) kernel radii from all 4 conditions.

### 3.3 Cinematic volume rendering

To demonstrate the cinematic volume rendering, we show two examples of depth-unaware and -aware composite volume rendering using functions of `create_composite_rendering_with_mask_associated_presets()` and `multi_volume_rendering_with_presets()` on bones, liver and cardiovascular group.

The rendering settings for the 3 different groups can be loaded from 3D slicer's MRML file via function `get_rendering_preset_from_xml()` from volRenderer module from AnatomyArchive by locating the rendering file "presets.xml" with full path and specifying `name_preset` to 'CT-Coronary-Arteries-2', 'CT-Liver-Vasculature' or 'CT-Cardiac3'.

Suppose that the segmentation results of NifTI image format from task "total" of the original image `img_nii` is `mask_nii`. The VTK image containing only anatomies belonging to the cardiovascular group can be created using codes below:

```
img_vtk = volRenderer.read_nifti_as_vtk_image(img_nii)
mask_array, affine, _ = genericImageIO.convert_nifti_to_numpy(mask_nii)
selected_labels=[k for k, v in get_v_dependent_cls_map('total').items() if v in
                 anatomy_groups['cardiovascular']]
list_mask = [mask_array==k for k in selected_labels]
mask_vtk = volRenderer.merge_list_masks_and_convert_to_vtk_image(list_mask, affine)
img_vtk_filtered = volRenderer.prepare_masked_volume_for_rendering(img_vtk,
                                                                  mask_vtk,
                                                                  reverse=False)
```

If the aim is to filter away the cardiovascular group as shown in Figure 7A to make the rendering of the skeleton clean, one can simply set the reverse tag in the text box above to True.

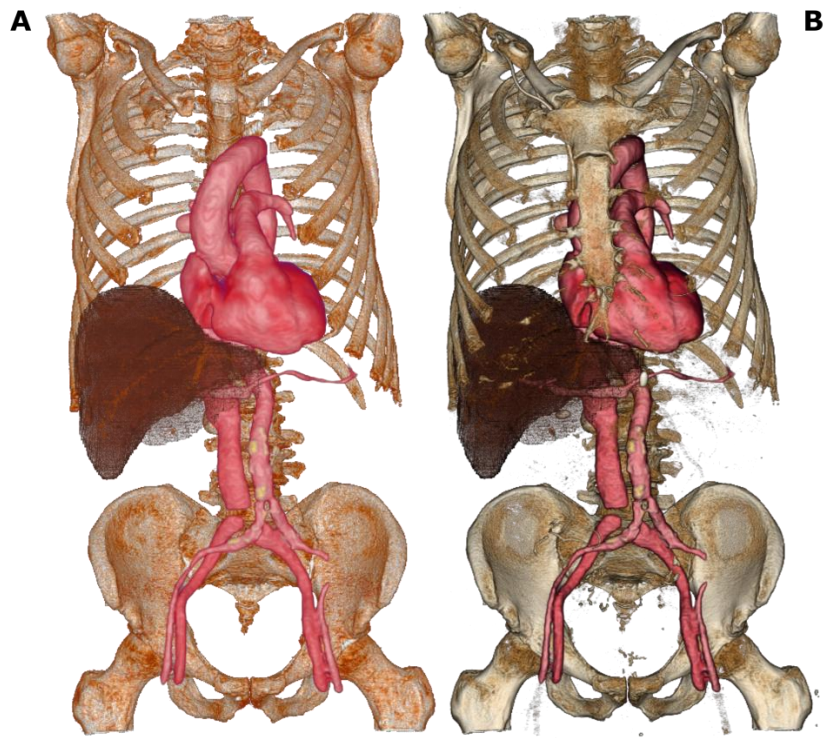

**Figure** 7. Composite rendering of segmented cardiovascular group and liver using depth-unaware renderer and depth-aware multi-volume renderer in presence of background bones (without applying their segmentation masks). Liver was specifically rendered with slight transparency to allow easier inspection of internal vasculature.